\documentclass[final]{siamltex}

\usepackage{latexsym}
\usepackage{amssymb}
\usepackage{amsfonts}
\usepackage{graphicx}
\usepackage{subfigure}
\usepackage{lineno}
\usepackage{subfigure}

\newtheorem{remark}[theorem]{Remark}

\title{Boundary driven waveguide arrays: Supratransmission and saddle-node bifurcation\thanks{Received by the editors \dots.}}

\author{
Hadi Susanto\thanks{School of Mathematical Sciences,
University of Nottingham, University Park, Nottingham, NG7 2RD, UK
({\tt hadi.susanto@nottingham.ac.uk})}}

\begin{document}

\maketitle

\begin{abstract}
In this report, we consider a semi-infinite discrete nonlinear Schr\"odinger equation driven at one edge by a driving force. The equation models the dynamics of coupled waveguide arrays. When the frequency of the forcing is in the allowed-band of the system, there will be a linear transmission of energy through the lattice. Yet, if the frequency is in the upper forbidden band, then there is a critical driving amplitude for a nonlinear tunneling, which is called supratransmission, of energy to occur. Here, we discuss mathematically the mechanism and the source of the supratransmission. By analyzing the existence and the stability of the rapidly decaying static discrete solitons of the system, we show rigorously that two of the static solitons emerge and disappear in a saddle-node bifurcation at a critical driving amplitude. One of the emerging solitons is always stable in its existence region and the other is always unstable. We argue that the critical amplitude for supratransmission is then the same as the critical driving amplitude of the saddle-node bifurcation. We consider as well the case of the forcing frequency in the lower forbidden band. It is discussed briefly that there is no supratransmission because in this case there is only one rapidly decaying static soliton that exists and is stable for any driving amplitude.
\end{abstract}



\pagestyle{myheadings}
\thispagestyle{plain}
\markboth{HADI SUSANTO}{BOUNDARY DRIVEN WAVEGUIDE ARRAYS}

\section{Introduction}
\label{intro}

An exotic nonlinear phenomenon has been discovered recently by Geniet and Leon \cite{geni02} in a semi-infinite chain of coupled oscillators driven at one edge by a time periodic forcing. Energy excitations will propagate through the chain if the driving frequency is in the allowed-band of the discrete system. It is natural because of the system's dispersion relation. In contrast, it would be expected that if the forcing frequency is in the band-gap, then there would be no energy flow. Yet, Geniet and Leon \cite{geni02} show theoretically and experimentally that there is a definite driving amplitude threshold above which a sudden energy flow takes place. This phenomenon is called nonlinear supratransmission \cite{geni02}. An exciting independent work on a modified Klein-Gordon equation describing the Josephson phase of layered high-T$_{\rm{c}}$ superconductors shows the presence of the same phenomenon \cite{save06}. Promising technological applications employing supratransmission have been proposed as well accompanying these findings, such as binary signal transmissions of information \cite{maci07} and terahertz frequency selection devices \cite{save07}.

In \cite{khom04} Khomeriki considers boundary driven coupled optical focusing waveguide arrays described by
\begin{eqnarray}
i\frac{\partial\psi_n}{\partial z}=-\psi_{n+1}-\psi_{n-1}-\gamma|\psi_{n}|^2\psi_{n},\quad\psi_0=Ae^{i\Delta z},
\label{eq1}
\end{eqnarray}
with $\gamma>0$ and $n=1,2,\dots$. Here, $\psi_n$ is the electromagnetic wave amplitude in the $n$th guiding core, $z$ is the propagation variable, $\Delta$ is the propagation constant or the driving frequency, and $\gamma$ represents the nonlinearity coefficient which is taken to be $\gamma=2$ in this report. This model can also be considered as a slow modulation wave approximation to the discrete sine-Gordon equation \cite{kivs92}. Similar to the nonlinear band-gap tunneling observed by Geniet and Leon \cite{geni02}, it is reported that there is a critical threshold $A_{\rm{th}}(\Delta)$ for supratransmission when the propagation constant $\Delta$ is in the forbidden band $\Delta>2$ \cite{khom04}.

\begin{figure*}[tb]
\begin{center}
\hspace{0.4\textwidth}
\hspace{0.4\textwidth}
\hspace{0.4\textwidth}
\hspace{0.4\textwidth}
\end{center}
\caption{Three dimensional plots of time evolution of the boundary driven waveguide arrays Eq.\ (\ref{eq1}). When the driving frequency $-2<\Delta<2$ is in the allowed band, any driving amplitude will lead to an energy flow to remote sites (top left panel). If $\Delta$ is in the upper forbidden band and $A$ is small enough, the boundary will excite a couple of arrays only (top right panel). Yet, there is a critical threshold amplitude $A_{\rm{th}}(\Delta)$ above which there is nonlinear forbidden band tunneling indicated by the released of a train of discrete solitons (bottom left panel). A quantitatively different behavior of supratransmission occurs when the driving frequency is large enough as is shown in the bottom right panel.}
\label{fig1}
\end{figure*}

In Fig.\ \ref{fig1}, we present numerical simulations of the dynamics of Eq.\ (\ref{eq1}). Following \cite{khom04}, the driving is turned on adiabatically to avoid the appearance of an initial shock by assuming the form
\[
A=\breve{A}(1-\exp(-z/\tau)),
\]
where we omit the breve henceforth. In the following figures, we take $\tau=50$ and apply a linearly increasing damping to the last 20 sites to suppress edge reflection.

Presented in the top left panel of Fig.\ \ref{fig1} is a three dimensional plot of time evolution of Eq.\ (\ref{eq1}) when the driving frequency is in the allowed band $-2<\Delta<2$. A small driving amplitude will excite all the sites. On the other hand, if the driving frequency is in the upper band $\Delta>2$, a small $A$ will only excite several neighboring sites as is shown in the top right panel of Fig.\ \ref{fig1}. Yet, if the driving amplitude is large enough, then a train of 'traveling' discrete solitons can be released \cite{khom04} (see bottom left panel of the same figure). This flow of energy is the so-called supratransmission or nonlinear forbidden band tunneling and we call the minimum $A$ for supratransmission to occur a critical threshold $A_{\rm{th}}$. The word \emph{traveling} is in quotes because Eq.\ (\ref{eq1}) does not admit a genuine one (see \cite{melv06} and references therein). If one waits long enough, the gap-solitons will be trapped by the lattice. Khomeriki \cite{khom04} also notices an immediate trapping when the driving frequency $\Delta$ is relatively large as is shown in the bottom right panel of Fig.\ \ref{fig1}. In this regime, the corresponding discrete gap solitons are highly localized.

An analytical approximation of $A_{\rm{th}}(\Delta)$ in the limit $0<\Delta-2\ll1$ is given by \cite{khom04}
\begin{equation}
A_{\rm{th}}(\Delta)=\sqrt{\Delta-2}.
\label{app_kh}
\end{equation}

\begin{remark}
Equation (\ref{eq1}) is symmetric with respect to the transformation $A\to-A$ and $\psi_n\to-\psi_n$. This means that there is also a critical amplitude $-A_{\rm{th}}(\Delta)$ if one applies $A<0$ such that for $A<-A_{\rm{th}}(\Delta)<0$, a nonlinear forbidden band tunneling will occur. Equation (\ref{eq1}) is also symmetric with respect to the transformation $\Delta\to-\Delta$, $\psi_n\to(-1)^n\psi_n$, and $\gamma\to-\gamma$. Therefore, the same phenomenon can be observed in defocusing waveguide arrays $\gamma<0$. Due to the transformation, the only difference of defocusing arrays from the self-focusing ones is that there will be a $\pi$ phase-difference between neighboring lattices.
\end{remark}

It is presented in \cite{khom04} that the numerical result for the threshold amplitude deviates rapidly from the approximation (\ref{app_kh}). It is because Eq.\ (\ref{app_kh}) is actually the amplitude-temporal frequency relation of the continuous nonlinear Schr\"odinger (NLS) equation's solitons. The relation has been phase-shifted properly due to some transformation, i.e.\ $\psi_n\to\psi_n\exp(2iz)$. Applying the transformation to (\ref{eq1}) will take it to a normalized standard finite difference approximation of the continuous NLS equation.

Remembering the aforementioned promising applications of nonlinear tunneling, it is therefore of interest to obtain an approximation of the threshold amplitude in the other limit $\Delta-2\gg1$. It is one of the aims of the present report. The other aim is to understand mathematically the mechanism of the nonlinear tunneling. It is mentioned, but not rigorously proved, \cite{khom04_2} that the supratransmission happens because of the emergence of two solutions at the critical driving amplitude, i.e.\ a saddle-node bifurcation. If it is the case, this then means that the threshold amplitude is not necessarily the amplitude of the corresponding fundamental soliton Eq.\ (\ref{app_kh}).  Understanding the source of supratransmission also will allow us to explain, e.g., the reason why there is no threshold amplitude for nonlinear tunneling when the driving frequency is in the lower forbidden band $\Delta<-2$.

Nonetheless, one may question the relevance of our first aim,
as supratransmission is quickly trapped by the lattices for large $\Delta$. Even though our analysis may be not immediately applicable to the present case, the aim is still of relevance. There are several experimentally realizable discrete equations that support 'traveling' solitons in a parameter region where the gap solitons are highly localized. As a particular example is the discrete Schr\"odinger equation with saturable nonlinearity in the large nonlinearity coefficient regime \cite{hadz04}. We have observed supratransmission in this equation and have successfully applied our analysis presented herein to obtain an approximation to the threshold amplitude \cite{susa08}. Later on in this paper, we also conjecture that our analytically obtained approximation, presented in terms of a power series expansion, may well be convergent uniformly in the region of interest, i.e., $\Delta>2$. Moreover, the mathematical procedure presented herein can also be applied as an alternative method to analyze the bistability effect considered, e.g., in \cite{khom05}. We might even consider it simpler and more appropriate as the analysis can then be done solely in its discrete set-up, with no necessity of approximating the problem 
with its continuous counterpart \cite{khom05}. 

In this study, we will show that the supratransmission is indeed related to saddle-node bifurcations. To mathematically prove this, our strategy is as follows. We will first prove the existence of a mode bifurcating from the constant solution $\psi_n\equiv0$ due to the driving site. 
We will also show that there is a singular mode bifurcating from infinity. We will then demonstrate that these two modes collide in a saddle-node bifurcation by developing an asymptotic analysis in the range of $\Delta$ large. Such an analysis is doable in that regime because the modes are highly localized. The final step to show that the critical amplitude is the same as the threshold amplitude for supratransmission is to prove that the mode bifurcating from zero state is stable, all the way on its existence region. Using this result, then we can derive an approximation of the threshold amplitude in terms of a power series expansion that can be calculated to any order. Numerical computations will be presented as well to compare our analytical results.

Our paper will be outlined in the following structures. In Subsection 2.1, we present our asymptotic analysis for the existence of monotonically decaying static solutions of Eq.\ (\ref{eq1}). The next subsection will contain our study on the stability analysis of solutions discussed in the preceding subsection. Using the same procedures, we then briefly discuss in Subsection 2.3 that there is no supratransmission in the case of $\Delta<-2$ as there is no bifurcation occurring in this regime. Then, we compare our analytical findings with the results of numerical computations in Subsection 2.4. Finally, we summarize our findings and present our conclusions in the last section.

\section{Existence and stability analysis of rapidly decaying discrete solitons}


\subsection{Existence analysis}

Stationary solutions of Eq.\ (\ref{eq1}) are sought in the form of $\psi_n(z)=\phi_ne^{i\Delta z}$, where $\phi_n$ is a real valued function. This ansatz can be applied as one would naturally expect that all the sites will be excited with the same frequency as the driving frequency. Since we are interested in the large propagation constant $\Delta$, we scale $\Delta\to1$ and define $\epsilon=1/\Delta$. Hence, we consider $|\epsilon|\ll1$. Static equation of Eq.\ (\ref{eq1}) is then given by
\begin{eqnarray}
F(\phi,\epsilon):=-\phi_n+\epsilon\left(\phi_{n+1}+\phi_{n-1}+\gamma{\phi_n}^3\right)=0,
\label{eq2}
\end{eqnarray}
with $\phi_0=A.$

When $|\epsilon|$ is small enough, apart from the boundary, the leading order solution of $\phi_n$ would formally satisfy
\begin{equation}
\phi_n\left(-1+\epsilon\gamma{\phi_n}^2\right)\approx0,
\label{tamb}
\end{equation}
from which we obtain that $\phi_n\approx0$ and $\phi_n\approx\pm1/\sqrt{\epsilon\gamma}$.
It physically means that the arrays are almost uncoupled and indicates that solutions of Eq.\ (\ref{eq2}) can be expressed in terms of an asymptotic or a perturbation expansion 
in $\epsilon$. 
It also says that when we consider finitely long waveguide arrays, i.e.\ $n=1,2,\dots,N$, Eq.\ (\ref{eq2}) can have at most $3^N$ solutions. Yet, only some of them are related to the nonlinear tunneling phenomenon presented in Fig.\ \ref{fig1}. We are especially interested in solutions with a magnitude that is monotonically decaying with the property $|\phi_n|\to0$ as $n\to\infty$. This consideration is based on the fact that when the driving frequency is in the forbidden band and the driving amplitude is below the critical threshold, the solution profile is monotonically decaying as $n\to\infty$ (see the top right panel of Fig.\ \ref{fig1}). Moreover, we only need to consider particularly a family of \emph{rapidly decaying discrete solitons} which is defined as follows.

\begin{definition}
Let $\phi_n=\sum_{k=0}^\infty a_{n,k}\vartheta_k(\epsilon)$ be a solution of (\ref{eq2}), where $n\in\mathbb{Z}^+,$ $\vartheta_k(\epsilon)$ is an asymptotic sequence and $\vartheta_k(\epsilon)=o\left(\vartheta_{k-1}(\epsilon)\right)$ for $\epsilon\to0$.
$\phi_n$ is a rapidly decaying discrete soliton of Eq.\ (\ref{eq2}) if $|\phi_n|$ is a monotonically decreasing function to 0 as $n\to\infty$ with a property that 
to the leading order $\mathcal{O}(\vartheta_0)$ only the first lattice site is non-zero, i.e.\ $a_{1,0}\neq0$ and $a_{n,0}=0,\,n\neq1$.
\end{definition}

As an example of this definition, let us consider the following solution of (\ref{eq2})
\begin{equation}
\Phi_0(n,A)=\left\{
\begin{array}{lll}
\displaystyle -\frac1{\sqrt{\gamma}}\left(\frac1{\sqrt\epsilon}+\frac{\sqrt\epsilon}{2}\right)-A\frac\epsilon2+\mathcal{O}(\epsilon^{3/2}),\,n=1\\
\displaystyle \frac1{\sqrt{\gamma}}\left(\frac1{\sqrt\epsilon}+\frac{\sqrt\epsilon}{2}\right)+\mathcal{O}(\epsilon^{3/2}),\,n=2\\
\displaystyle \mathcal{O}(\sqrt\epsilon),\,\textrm{otherwise}.
\end{array}
\right.
\label{tamb2}
\end{equation}

This solution is obtained from the expansion: 
$\phi_1=-1/\sqrt{\epsilon\gamma}+a_{1,1}\sqrt\epsilon+a_{1,2}\epsilon+\dots$, $\phi_2=1/\sqrt{\epsilon\gamma}+a_{2,1}\sqrt\epsilon+a_{2,2}\epsilon+\dots$, $\phi_3=0+a_{3,1}\sqrt\epsilon+\dots$, and $\phi_n=0+\dots$ for $n>3$. Substituting the ansatz to Eq.\ (\ref{eq2}) will yield polynomials in $\epsilon$. Equating the coefficients of the polynomials for all orders of $\epsilon$ to zero will yield equations for $a_{k,l}$ that have to be solved simultaneously to obtain Eq.\ (\ref{tamb2}).

It is clear that the profile of $|\Phi_0(n,A)|$ (\ref{tamb2}) is monotonically decaying in $n$. However, this solution is not rapidly decaying as to the leading order, i.e.\ $\mathcal{O}(1/\sqrt\epsilon)$, $|\Phi_0(2,A)|=|\Phi_0(1,A)|\neq0$.

The existence of rapidly decaying solutions of (\ref{eq2}) when $A=\mathcal{O}(1)$ is guaranteed by the following theorem.

\begin{theorem}
Let $A$ be of $\mathcal{O}(1)$. Then for $\epsilon$ positive and small there are three rapidly decaying discrete solitons of the static equation (\ref{eq2}). Denoted by $\Phi_j$, $j=1,2,3$, the solitons are given by
\begin{eqnarray}
\Phi_1(n,A)&=&\left\{
\begin{array}{lll}
\displaystyle \frac1{\sqrt{\epsilon\gamma}}-A\frac\epsilon2-\frac{\epsilon^{3/2}}{2\sqrt\gamma}+\mathcal{O}(\epsilon^2),\,n=1\\
\displaystyle \frac{\sqrt\epsilon}{\sqrt\gamma}+\mathcal{O}(\epsilon^2),\,n=2\\
\displaystyle \frac{\epsilon^{3/2}}{\sqrt\gamma}+\mathcal{O}(\epsilon^2),\,n=3\\
\displaystyle 0+\mathcal{O}(\epsilon^2),\,\textrm{otherwise},
\end{array}
\right.\label{Phi1}\\
\Phi_2(n,A)&=&\left\{
\begin{array}{lll}
\displaystyle A\epsilon+A\epsilon^3+\mathcal{O}(\epsilon^5),\,n=1\\
\displaystyle A\epsilon^2+\mathcal{O}(\epsilon^4),\,n=2\\
\displaystyle A\epsilon^3+\mathcal{O}(\epsilon^5),\,n=3\\
\displaystyle 0+\mathcal{O}(\epsilon^4),\,\textrm{otherwise},
\end{array}
\right.\label{Phi2}\\
\Phi_3(n,A)&=&\left\{
\begin{array}{lll}
\displaystyle -\frac1{\sqrt{\epsilon\gamma}}-A\frac\epsilon2+\frac{\epsilon^{3/2}}{2\sqrt\gamma}+\mathcal{O}(\epsilon^2),\,n=1\\
\displaystyle -\frac{\sqrt\epsilon}{\sqrt\gamma}+\mathcal{O}(\epsilon^2),\,n=2\\
\displaystyle -\frac{\epsilon^{3/2}}{\sqrt\gamma}+\mathcal{O}(\epsilon^2),\,n=3\\
\displaystyle 0+\mathcal{O}(\epsilon^2),\,\textrm{otherwise}.
\end{array}
\right.\label{Phi3}
\end{eqnarray}

\label{e_sA}
\end{theorem}

\begin{proof}
Because we are looking for rapidly decaying solitons, to the leading order Eq.\ (\ref{eq2}) can be represented by
\begin{equation}
-\phi_1+\epsilon A+\gamma\epsilon{\phi_1}^3=0.
\label{lo}
\end{equation}
Equation (\ref{lo}) is a cubic equation similar to Eq.\ (\ref{tamb}), also with three roots. However, as $\epsilon\to0$, (\ref{lo}) reduces to a linear equation $\phi_1=0$ with only a single root. Therefore, finding the roots of the equation is a singular perturbation problem. Following, e.g., \cite{nayf81} (see Example 3 of Section 2.1 and 2.2), one will obtain the roots of (\ref{lo}), i.e.\ $\phi_1=A\epsilon+\dots$ and $\phi_1=\pm1/\sqrt{\gamma\epsilon}+\dots$. This concludes that there are three rapidly decaying solutions of (\ref{eq2}). In the following, let us name the solitons $\Phi_j(n,A)$, $j=1,2,3,$ with $\Phi_1(1,A)=1/\sqrt{\epsilon\gamma}+\dots$, $\Phi_2(1,A)=\epsilon A+\dots,$ and $\Phi_3(1,A)=-1/\sqrt{\epsilon\gamma}+\dots$.
%
The existence of $\Phi_j(n,A)$ for Eq.\ (\ref{eq2}) follows immediately from the Implicit Function Theorem (see, e.g., \cite{nire74}) since $F$ is differentiable and the Jacobian matrix of problem (\ref{eq2}) $DF(\phi,0)$ is invertible. Explicit calculations to obtain (\ref{Phi1})-(\ref{Phi3}) can be done similarly following the derivation of (\ref{tamb2}). 
%
\end{proof}

If one compares the above theorem and the top right panel of Fig.\ \ref{fig1}, it can be recognized immediately that the solution observed in the panel in the limit $z\to\infty$ is nothing else but $|\Phi_2(n,A)|$.


One still can obtain the existence of the above rapidly decaying solutions even when $A\gg1$ as stated in the following theorem.

\begin{theorem}
Let $A$ be scaled to $A=\tilde{A}/\epsilon^{3/2}$, $\tilde{A}<2/\sqrt{27\gamma}$.
\begin{equation}
\Phi_j(n,A)=\left\{
\begin{array}{lll}
\displaystyle \frac{\Phi_j^0}{\sqrt\epsilon}+\frac{\tilde{A}}{3\gamma{\Phi_j^0}^2-1}\left(\sqrt\epsilon-\epsilon\right)+\mathcal{O}(\epsilon^{3/2}),\,n=1\\
\displaystyle \Phi_j^0\sqrt\epsilon+\mathcal{O}(\epsilon^{3/2}),\,n=2\\
\displaystyle 0+\mathcal{O}(\epsilon^{3/2}),\,\textrm{otherwise},
\end{array}
\right.
\label{PhijAg}
\end{equation}
with $\Phi_j^0$ given by
\begin{equation}
\Phi_j^0=
\left\{
	\begin{array}{lll}
		\displaystyle \frac{2}{\sqrt{3\gamma}} \cos \left(\frac{1}{3}\arccos\left( \frac{-\tilde{A}\sqrt{27\gamma}}{2} \right) \right)
		,\quad j=1,\\
		\displaystyle \frac{2}{\sqrt{3\gamma}} \cos \left(\frac{4\pi}{3}+\frac{1}{3}\arccos\left( \frac{-\tilde{A}\sqrt{27\gamma}}{2} \right) \right)
		,\quad j=2,\\
		\displaystyle \frac{2}{\sqrt{3\gamma}} \cos \left(\frac{2\pi}{3}+\frac{1}{3}\arccos\left( \frac{-\tilde{A}\sqrt{27\gamma}}{2} \right) \right)
		,\quad j=3.
	\end{array}\right.
	\label{Phij0}
\end{equation}
Moreover, if we write $A=2/\sqrt{\left(27\gamma\epsilon^3\right)}-\widehat{A}\sqrt\epsilon$, with $\widehat{A}>1/\sqrt{3\gamma}$, then $\Phi_j$, $j=1,2$, can be written as
\begin{eqnarray}
\Phi_{1,2}&=&\left\{
\begin{array}{lll}
\displaystyle 
\frac1{\sqrt{3\gamma}\sqrt\epsilon}\mp\sqrt{\frac{\widehat{A}}{\sqrt{3\gamma}}-\frac1{3\gamma}}\sqrt\epsilon+\mathcal{O}(\epsilon^{3/2}),\,n=1\\
\displaystyle \frac{\sqrt\epsilon}{\sqrt{3\gamma}}+\mathcal{O}(\epsilon^{3/2}),\,n=2\\
\displaystyle \mathcal{O}(\epsilon^{3/2}),\,\textrm{otherwise}.
\end{array}
\right.\label{Phi12b}
\end{eqnarray}
\label{e_bA}
\end{theorem}

\begin{proof}
As we are interested in the case of $A\gg1$, we first scale
$A=\tilde{A}/\epsilon^{3/2}$ and correspondingly write $\Phi_j(n,A)=\Phi_j^0(n,A)/\sqrt\epsilon+\dots
$,
$j=1,2,3$, with $\Phi_j^0(n,A)=0$ for all $1<n\in\mathbb{Z}^+$. Substituting the expansion to Eq.\ (\ref{eq2}) and identifying coefficients for power series of $\mathcal{O}(1/\sqrt\epsilon)$ yields the following cubic equation for $\Phi_j^0(1,A)=\Phi_j^0$, i.e.\
\begin{equation}
G\left(\Phi_j^0\right):=-\Phi_j^0+\tilde{A}+\gamma\left(\Phi_j^0\right)^3=0.
\label{Phi}
\end{equation}

Equation (\ref{Phi}) cannot be solved perturbatively to obtain the roots $\Phi_j^0$ as before as all the terms in (\ref{Phi}) are of the same order. Therefore, we need the following lemma on cubic equations.
\begin{lemma}
Consider the following polynomial equation
\[
g(x)=ax^3+bx^2+cx+d,\quad a,b,c,d\in\mathbb{R}.
\]
Let
\begin{eqnarray}
\displaystyle X=\frac{-b}{3a},\quad Y=g(X),\quad h=2a\upsilon^3,\nonumber\\
\displaystyle \upsilon^2=\frac{b^2-3ac}{9a^2},\quad \theta=\frac13\arccos(\frac{-Y}{h}).\nonumber
\end{eqnarray}
If 
$Y^2<h^2$,
then the cubic equation has three distinct real roots given by
\begin{eqnarray}
x_1&=&X+2\upsilon\cos\theta
,\label{wi1}\\
x_2&=&X+2\upsilon\cos(4\pi/3+\theta)
,\label{wi2}\\
x_3&=&X+2\upsilon\cos(2\pi/3+\theta)
,\label{wi3}
\end{eqnarray}
where
\[
x_1> x_2> x_3.
\]
When $Y^2=h^2$, two of the roots which are neighboring to each other, i.e.\ $x_1$ and $x_2$ or $x_2$ and $x_3$, will collide in a saddle-node bifurcation and disappear when $Y^2>h^2$, i.e.\ the cubic equation then only has a single real root.
\label{cp}
\end{lemma}

\begin{proof}
See \cite{wiki}. The expression of the cubic polynomial roots (\ref{wi1})-(\ref{wi3}) is using Nickalls' geometric representation \cite{nick93}. 
\end{proof}

According to Lemma \ref{cp}, 
Eq.\ (\ref{Phi}) has geometric representation parameters:
\begin{eqnarray}
\displaystyle X=0,\quad Y=\tilde{A},\quad \upsilon=\frac{1}{\sqrt{3\gamma}},\quad h=\frac{2}{\sqrt{27\gamma}},\quad\theta=\frac13\arccos(\frac{-Y}{h}),\nonumber
\end{eqnarray}
from which we can conclude that (\ref{Phi}) has three real roots when $\tilde{A}<2/\sqrt{27\gamma}$. The roots of (\ref{Phi}), i.e.\ Eqs.\ (\ref{Phij0}), are obtained using Eqs.\ (\ref{wi1})-(\ref{wi2}). Then, the continuation of $\Phi_j^0$ can be obtained immediately using the Implicit Function Theorem.


It is then straightforward to calculate that when $\tilde{A}=2/\sqrt{27\gamma}$, $\Phi_{1,2}^0=1/\sqrt{3\gamma}$ as $2\cos\theta=2\cos(4\pi/3+\theta)=1$. Hence, we know that $\Phi_1(n,A)$ collides with $\Phi_2(n,A)$ in a saddle-node bifurcation.

For the value of $A$ close to the occurrence of the saddle-node bifurcation, we write $A=2/\sqrt{27\gamma\epsilon^3}-\widehat{A}\sqrt\epsilon$. In this case, the Implicit Function Theorem cannot be immediately employed to prove the existence of $\Phi_1$ and $\Phi_2$ as we need a bound for $\widehat{A}$. 

First, we substitute to the steady state equation (\ref{eq2}) $\Phi_{j}=\Phi_{j}^0(n,A)/\sqrt\epsilon+\sqrt\epsilon\Phi_{j}^1(n,A)$, $j=1,2$, with $\Phi_{j}^0(n,A)=1/\sqrt{3\gamma}$ for $n=1$ and $0$ otherwise. This then gives the following equations:
\begin{eqnarray}
\tilde{G}_1(\Phi^1_j,\epsilon)&:=&\Phi_{j}^1(2,A)-\widehat{A}+\sqrt{3\gamma}\left(\Phi_{j}^1(1,A)\right)^2+\epsilon\gamma\left(\Phi_{j}^1(1,A)\right)^3=0,\nonumber\\
\tilde{G}_2(\Phi^1_j,\epsilon)&:=&-\Phi_{j}^1(2,A)+\epsilon\Phi_{j}^1(3,A)+\frac1{\sqrt{3\gamma}}+\epsilon\Phi_{j}^1(1,A)+\epsilon^2\gamma\left(\Phi_{j}^1(1,A)\right)^3=0,\nonumber\\
\tilde{G}_n(\Phi^1_j,\epsilon)&:=&\Phi_{j}^1(n,A)+\epsilon\left(\Phi_{j}^1(n+1,A)+\Phi_{j}^1(n-1,A)+\epsilon\gamma\Phi_{j}^1(n,A)^3\right)=0,\quad n\neq1,2.\nonumber
\end{eqnarray}

Taking $\epsilon=0$, the above equations give us
\begin{eqnarray}
\Phi_{j}^1(1,A)&=&\pm\sqrt{\frac{\widehat{A}}{\sqrt{3\gamma}}-\frac1{{3\gamma}}},\nonumber\\
\Phi_{j}^1(2,A)&=&\frac1{\sqrt{3\gamma}},\nonumber\\
\Phi_{j}^1(n,A)&=&0,\quad n\neq1,2.\nonumber
\end{eqnarray}
Note that the $\pm$-solutions collide for $\widehat{A}=1/\sqrt{3\gamma}$. Because the linearization $D\tilde{G}(\Phi_j^1,0)$ is invertible for $\widehat{A}>1/\sqrt{3\gamma}$, the Implicit Function Theorem can be applied again and we have the existence of rapidly decaying solitons $\Phi_{j}=\Phi_{j}^0(n,A)/\sqrt\epsilon+\sqrt\epsilon\Phi_{j}^1(n,A)$, $j=1,2$.
\end{proof}


With this theorem, we then have shown that $\Phi_1$ collides in a saddle-node bifurcation with $\Phi_2$. Yet, we cannot directly claim that this is the source of the supratransmission observed in Fig.\ \ref{fig1} before we show and discuss the stability of the two solitons.

\subsection{Stability analysis}

After discussing the existence of rapidly decaying solitons of Eq.\ (\ref{eq2}), next we study their stability. If $\phi_n$, $n=1,2,\dots$, is a solution of (\ref{eq2}), then the linear spectral stability of $\phi_n$ can be obtained by substituting the ansatz $\psi_n=(\phi_n+\delta[v_n e^{i\lambda z}+\overline{w_n} e^{-i\overline{\lambda}z}])e^{i\Delta z}$ with $\lambda\in\mathbb{C},\,(v_n,w_n)\in\mathbb{C}^2,$ and $n\in\mathbb{Z}^+$ into Eq.\ (\ref{eq1}). Linearizing the equation to $\mathcal{O}(\delta)$, we obtain the following eigenvalue problem
\begin{equation}
		\lambda\epsilon\left(\begin{array}{cc}v_n\\ w_n\end{array}\right) =
		\epsilon\sigma\left(\begin{array}{cc}v_{n-1}\\ w_{n-1}\end{array}\right) +
		\mathcal{L}\left(\begin{array}{cc}v_{n}\\ w_{n}\end{array}\right)+
				\epsilon\sigma\left(\begin{array}{cc}v_{n+1}\\ w_{n+1}\end{array}\right),
\label{eig}
\end{equation}
with
\begin{eqnarray}
&&\left(\begin{array}{c} v_0\\w_0\end{array} \right)=\left(\begin{array}{c} 0\\0\end{array} \right), \quad \sigma=
\left(
\begin{array}{cc}
1 & 0 \\
0 & -1
\end{array} \right),
\nonumber\\
&& \mathcal{L}=
\left(
\begin{array}{cc}
-1+2\epsilon\gamma|\phi_n|^2 & \epsilon\gamma{\phi_n}^2 \\
-\epsilon\gamma{\phi_n}^2 & 1-2\epsilon\gamma|\phi_n|^2
\end{array} \right), \quad n\in\mathbb{Z}^+,\nonumber
\end{eqnarray}
where we have scaled $\Delta\to1$.

The natural domain for $\tilde{\mathcal{L}}=\left( \epsilon\sigma\quad\mathcal{L}\quad\epsilon\sigma \right)$ is $L^2(\mathbb{C})$. We call $\lambda$ an eigenvalue of $\tilde{\mathcal{L}}$ if there is a function $\{v_n\}_{n\in\mathbb{Z}^+},\,\{w_n\}_{n\in\mathbb{Z}^+}\in L^2(\mathbb{C})$ which satisfies (\ref{eig}). Since $\tilde{\mathcal{L}}$ depends smoothly on $A$, the eigenvalues of $\tilde{\mathcal{L}}$ will depend smoothly on $A$, too. $\phi_n$ is linearly stable if the imaginary part of $\lambda$ is zero, i.e.\ Im$(\lambda)=0$.

The continuous spectrum is obtained by substituting
\[
v_n = Ae^{ikn},\quad w_n=Be^{ikn},\quad \phi_n=0,
\]
to Eq.\ (\ref{eig}) from which we will obtain
\[
\epsilon\lambda=\pm2\epsilon\cos k\mp1.
\]
Thus, the continuous spectrum of solutions under investigation is the range
\begin{equation}
\lambda\in\left(-\frac1\epsilon-2,-\frac1\epsilon+2\right)\, \rm{and}\, \lambda\in\left(\frac1\epsilon-2,\frac1\epsilon+2\right).
\label{cont}
\end{equation}
As the continuous spectrum lies in the real axis, the stability of the solutions is only determined by the discrete spectrum, i.e.\ eigenvalues. For the solutions given in Theorems \ref{e_sA} and \ref{e_bA}, we have the following stability results.

\begin{theorem}
For small driving amplitude $A=\mathcal{O}(1)$, the various rapidly decaying discrete solitons have the following properties:
\begin{enumerate}
\item the discrete soliton $\Phi_1$ is unstable. It has a single imaginary eigenvalue.
\item the soliton $\Phi_2$ is strictly stable as the soliton has no discrete eigenvalues.
\item the discrete soliton $\Phi_3$ is stable. It has a single real eigenvalue.
\end{enumerate}
\label{s_sA}
\end{theorem}

\begin{proof}
We are looking for eigenvectors that are also rapidly decaying. Therefore, the eigenvalue problem Eq.\ (\ref{eig}) to the leading order can be approximated by the linear eigenvalue problem
\[
\lambda\epsilon\left(\begin{array}{cc}v_1\\ w_1\end{array}\right) = \mathcal{L}\left(\begin{array}{cc}v_1\\ w_1\end{array}\right),
\]
which gives the following approximate eigenvalues
\begin{equation}
\lambda=\pm\frac1\epsilon\sqrt{{3\left(\epsilon\gamma{\phi_n}^2\right)^2-4{\epsilon\gamma{\phi_n}^2}+1}}.
\label{lineig}
\end{equation}
In the above expression, we have taken into account the fact that $\phi_n\in\mathbb{R}$.

For the stability of $\Phi_1$ and $\Phi_3$, substitute $\phi_1=\Phi_j(1,A)$, $j=1,3$ into Eq.\ (\ref{lineig}). Taking the series expansion of the expression gives the following eigenvalue $\lambda$ for $\Phi_j(n,A)$, i.e.
\begin{equation}
\lambda=\left\{
\begin{array}{ll}
	{\epsilon^{-1/4}}\sqrt{2A\sqrt\gamma}i+\mathcal{O}(\epsilon^{1/4}),\quad j=1,\\
	{\epsilon^{-1/4}}\sqrt{2A\sqrt\gamma}+\mathcal{O}(\epsilon^{1/4}),\quad j=3.
\end{array}
\right.
\label{sta1}
\end{equation}
Because the eigenvalue of $\Phi_1(n,A)$ has a non-zero imaginary part, we conclude that to the leading order $\Phi_1$ is unstable, as opposed to $\Phi_3$.

As for $\phi_1=\Phi_2(1,A)$, the series expansion of Eq.\ (\ref{lineig}) gives
\begin{equation}
\lambda=1/\epsilon+\mathcal{O}(\epsilon^2).
\label{sta2}
\end{equation}
Because $\lambda$ is inside the continuous spectrum (\ref{cont}), then our assumption that the eigenfunction is rapidly decaying is not justified. Nonetheless, we know that $\Phi_2$ bifurcates from a uniform solution $\phi_n\equiv0$ which is stable. Because $L$ depends smoothly on $A$, we then can conclude that $\Phi_1$ has no eigenvalue.
\end{proof}

When the driving amplitude is large, we also have the following theorem
\begin{theorem}
For large driving amplitude $A=\tilde{A}\epsilon^{-3/2}$, the various rapidly decaying discrete solitons have the following properties:
\begin{enumerate}
\item the discrete soliton $\Phi_1$ is unstable with a single imaginary eigenvalue.
\item the soliton $\Phi_2$ is strictly stable  with a single real eigenvalue.
\item the discrete soliton $\Phi_3$ is in general stable with a single eigenvalue, except in a finite interval where our asymptotic analysis is inconclusive.
\end{enumerate}
To the leading order, the eigenvalue of the three solitons is given by
\begin{equation}
\lambda=K/\epsilon+\frac{\tilde{A}}{\Phi_j^0\left(3\gamma\left(\Phi^0_j\right)^2-1\right)}\left( \frac{\left(3\gamma^2\left(\Phi^0_j\right)^4-1\right)}{K}+K \right)+\mathcal{O}(\sqrt\epsilon),
\label{s_Phi0}
\end{equation}
with $K=\sqrt{3\left(\gamma{\Phi^0_j}^2\right)^2-4\gamma{\Phi_j^0}^2+1}.$ Moreover, by writing $A=2/\sqrt{27\gamma e^3}-\widehat{A}\sqrt\epsilon,$ the eigenvalue of $\Phi_{1,2}$ is given by
\begin{equation}
\lambda=\frac2{{3}^{1/4}\sqrt\epsilon}\sqrt{\mp\sqrt{\widehat{A}\sqrt{\frac\gamma3}-\frac13}}+\mathcal{O}(\sqrt\epsilon),
\label{s_Phi12}
\end{equation}
with the 'minus' sign for the eigenvalue of $\Phi_1$ and the 'plus' sign for $\Phi_2$.
\label{s_bA}
\end{theorem}

\begin{proof}
The proof of Theorem \ref{s_bA} is similar to the proof of Theorem \ref{s_sA}. The stability result of $\Phi_3$ cannot be deduced immediately because the expression of $\Phi_3$ is not trivial. The presence of a finite interval where our asymptotic analysis is inconclusive cannot be seen clearly. It is inconclusive because there is a range of $A$ in which $\lambda$ is in the domain of the continuous spectrum (\ref{cont}). A numerical proof will be presented in the following section.
\end{proof}

\subsection{Analysis for the case of $\Delta<-2$}

We omit the details and the rigorous proof, but it can be shown that for $\Delta<-2$, there is only one rapidly decaying soliton which is stable for any driving amplitude. The idea is as follows.

Instead of Eq.\ (\ref{eq2}), consider
\begin{eqnarray}
\phi_n=-\epsilon(\phi_{n+1}+\phi_{n-1})-\gamma\epsilon{\phi_n}^3,\quad\phi_0=A,
\label{eq2c}
\end{eqnarray}
where we again have scaled $0>\Delta\to1$ and define $\epsilon=1/|\Delta|$. For a rapidly decaying solution, the leading order equation of (\ref{eq2c}) is then given by
\begin{equation}
f:=\phi_1+\epsilon\left(A+\gamma{\phi_1}^3\right)=0.
\label{eq2d}
\end{equation}
It is clear that $f\to\pm\infty$ as $\phi\to\pm\infty$. Yet, $f$ has no critical point, i.e.\ $df/d\phi_1>0$. Therefore, one can conclude that $f$ is a monotonically increasing function which intersects the horizontal axis once, i.e.\ $f$ has one real root. The stability of this rapidly decaying solution might be determined immediately following Theorem \ref{s_sA} and \ref{s_bA}. Our numerics, which are not presented here, show that when $A=\mathcal{O}(1)$ the solution is stable with no discrete spectrum and when $A=\mathcal{O}(1/\epsilon^{3/2})$ there is an eigenvalue bifurcating from the upper edge of the continuous spectrum. Hence, the soliton is stable all the way to $A\to\infty$ which explains why there is no supratransmission for $\Delta<-2$.

\subsection{Numerical results}

To accompany our analytical results, we have used numerical calculations. For that purpose, we have made a continuation program based on a Newton iteration technique to obtain stationary rapidly decaying discrete solitons of Eq.\ (\ref{eq2}) and an eigenvalue problem solver to solve (\ref{eig}) in \textsc{MATLAB}. Throughout the subsection, we consider in particular 
$\Delta=10$. Even though there is no prominent supratransmission of energy for this value of $\Delta$, it is taken solely as an example to show that especially in the regime of $\Delta$ large, our asymptotic analysis explains the problem well. It will be shown below that, e.g., even using the first two terms of the approximate threshold amplitude, our analytical result is already relatively in agreement with the numerical results.

\begin{figure}[tbh]
\begin{center}
\includegraphics[width=.4\textwidth]{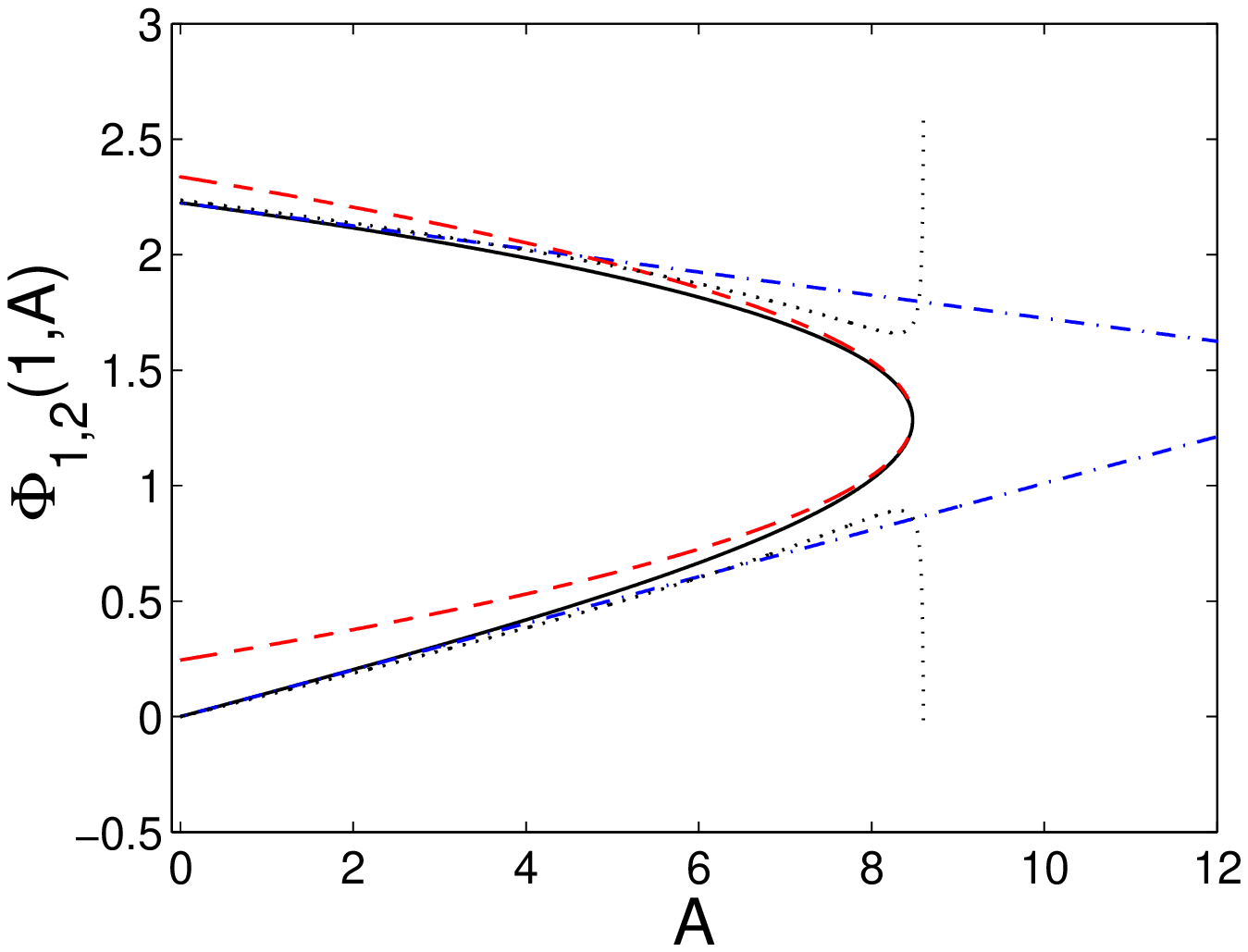}
\includegraphics[width=.4\textwidth]{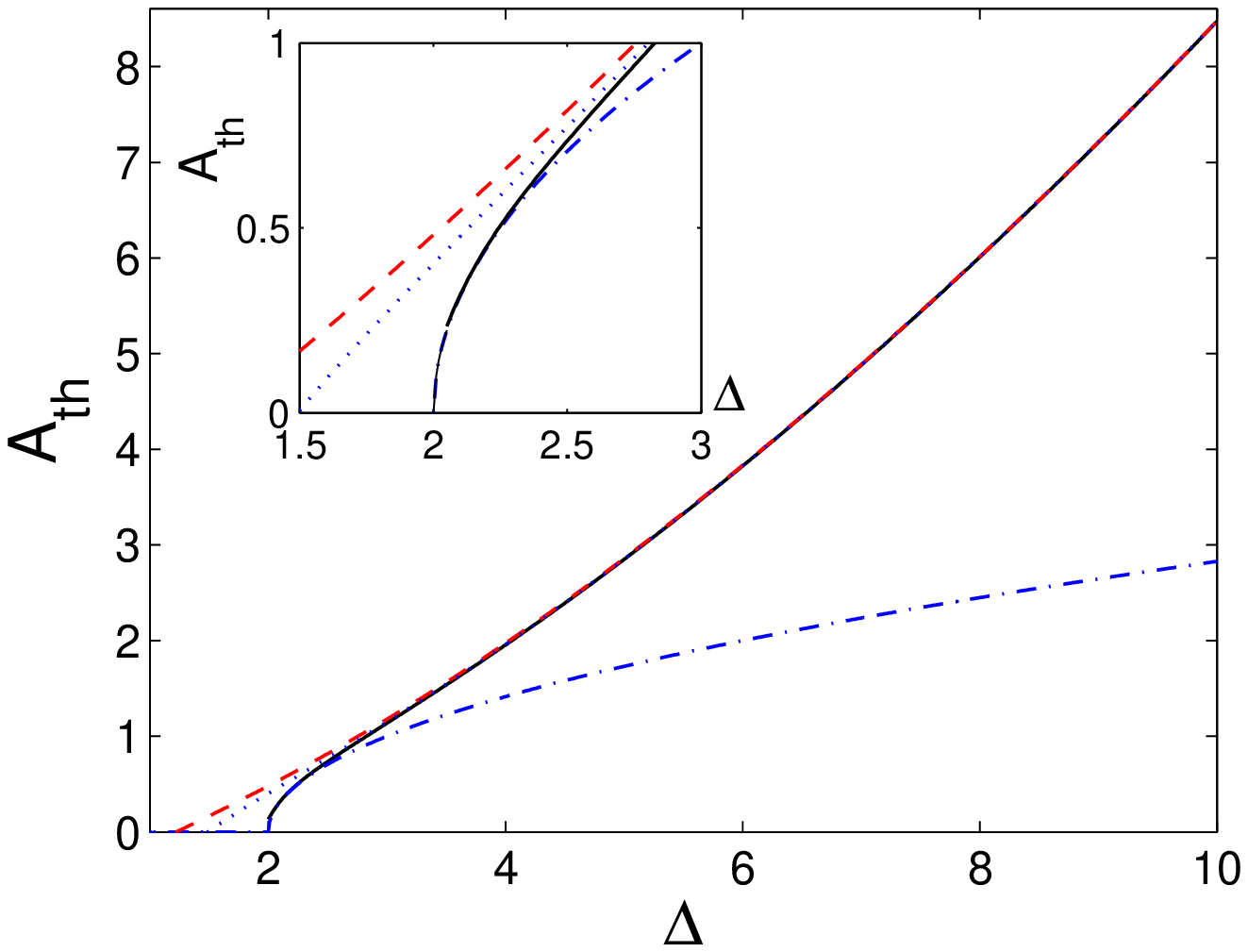}\\
\includegraphics[width=.4\textwidth]{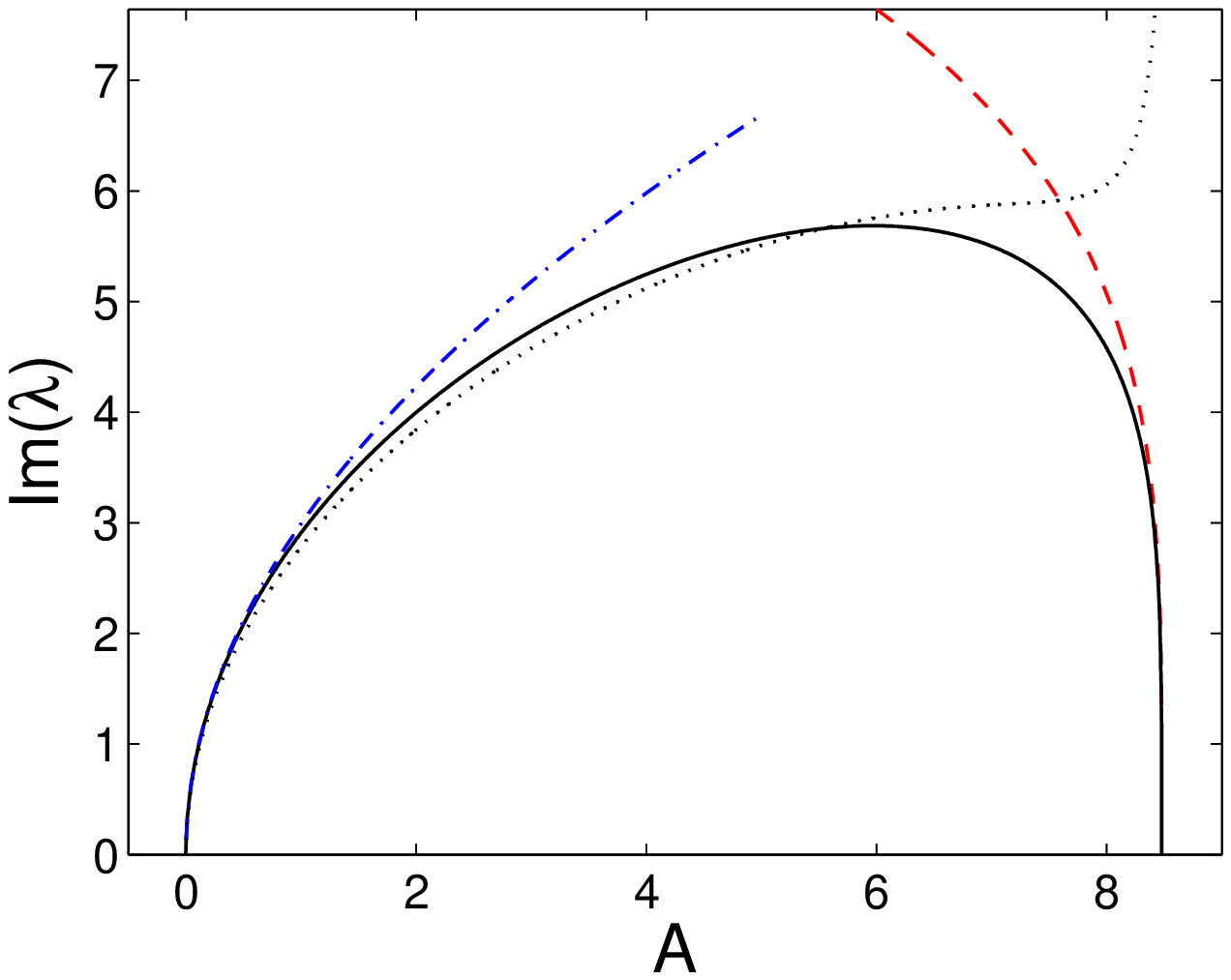}
\includegraphics[width=.4\textwidth]{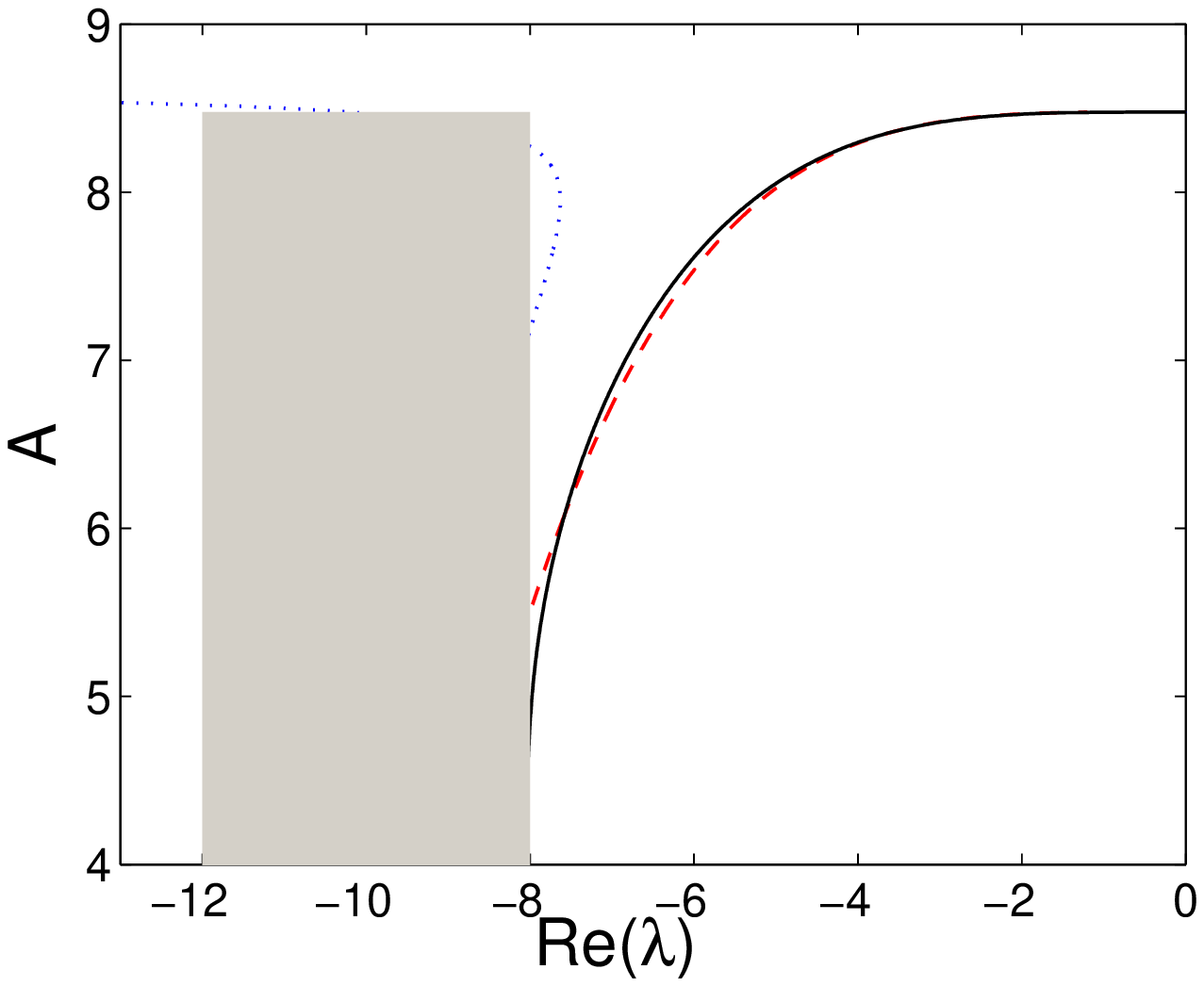}
\end{center}
\caption{Presented is the comparison between the numerically obtained results and the analytical calculations presented in Section 2. Top left panel: the existence curve of $\Phi_1$ and $\Phi_2$ represented by the solution of the first site, where the upper and lower branch corresponds to the existence curve of $\Phi_1$ and $\Phi_2$, respectively. Top right panel: the threshold amplitude $A_{\rm{th}}$ as a function of the propagation constant $\Delta$. Bottom panels: the critical eigenvalue of $\Phi_1$ (left) and $\Phi_2$ (right) as a function of the driving amplitude $A$. Shaded region in the bottom right panel shows the region for the continuous spectrum. Analytical approximations calculated in Section 2 are also presented as dashed, dotted, and dash-dotted lines (see the text).}
\label{Ath}
\end{figure}

We summarize our results and discussions for the existence and the stability of $\Phi_1$ and $\Phi_2$ in Fig.\ \ref{Ath}. At the top left panel of the figure, we present the existence of $\Phi_j$, $j=1,2$, represented by the solution of the first site, where the upper and lower branch corresponds to the existence curve of $\Phi_1$ and $\Phi_2$, respectively. Presented in solid line is the numerical results. Our analytical result Eqs.\ (\ref{Phi1}) and (\ref{Phi2}) which is supposed to be valid when $A=\mathcal{O}(1)$ is depicted as dash-dotted line. As for the analytical approximations for $A=\mathcal{O}(1/\epsilon^{3/2})$, i.e.\ Eqs.\ (\ref{PhijAg}) and (\ref{Phi12b}), they are presented as dotted and dashed line, respectively. It is interesting to note that Fig.\ \ref{Ath} shows clearly a good agreement between our analytical and the numerical results.

Top right panel of Fig.\ \ref{Ath} presents the comparison between the critical amplitude $A_{\rm{th}}(\Delta)$ calculated numerically from Eq.\ (\ref{eq2}) and our approximation $A_{\rm{th}}(\Delta)=2/\sqrt{27\gamma\epsilon^{3/2}}-\sqrt\epsilon/\sqrt{3\gamma}$ (see Theorem \ref{e_bA}) which are presented in solid and dashed line, respectively. The numerical results were also checked against the full dynamics of the original problem Eq.\ (\ref{eq1}), where an agreement is obtained as it should be provided that $\tau$ is large enough. Note the good agreement when $\Delta\gg1$. As a comparison with the analytical approximation obtained by Khomeriki \cite{khom04}, we also present $A_{\rm{th}}(\Delta)=\sqrt{\Delta-2}$ in dash-dotted line.

It is interesting to note that in the limit $\Delta\to2$ our analytical approximation does not diverge. As is shown in the inset of the top right panel figure, the difference of the approximate value of the threshold amplitude and the numerical result at $\Delta=2$ is about $50\%$. Using the same method presented in the preceding sections, we obtain that the first three terms of the approximation of $A_{\rm{th}}(\Delta)$ are actually given by
\begin{equation}
A_{\rm{th}}(\Delta)=\frac2{\sqrt{27\gamma\epsilon^{3/2}}}-\frac{\sqrt\epsilon}{\sqrt{3\gamma}}-
\frac{13\sqrt3}{36\sqrt\gamma}\epsilon^{5/2}.
\label{Ath2}
\end{equation}
The plot of this curve is depicted in the same panel as dotted line where one can see that the difference now has decreased by about $10\%$. This then motivates us to question whether the infinite power series of the approximate threshold amplitude $A_{\rm{th}}(\Delta)$ is actually convergent uniformly to the critical amplitude curve. Considering the fact that the region of interest is on $0<\epsilon\leq1/2$ and the coefficients of the power series are so far bounded, the answer might well be affirmative. Yet, this question is out of the scope of the present report and it is therefore addressed for future investigations.

After presenting the numerical and the analytical results for the existence of $\Phi_1$ and $\Phi_3$, next we consider the stability of the solitons. Bottom panels of Fig.\ \ref{Ath} present the comparison between the results. The left panel shows imaginary part of the critical eigenvalue of $\Phi_1$ as a function of $A$ in its existence region. It is clear that the soliton is always unstable. The right panel presents the eigenvalue of $\Phi_2$ as a function of the driving amplitude where one can see that the soliton is always stable, as opposed to $\Phi_1$. Our analytical approximations (\ref{sta1}), (\ref{s_Phi0}), and (\ref{s_Phi12}) are presented as well in the two panels as dash-dotted, dotted, and dashed line, respectively. It is also interesting to note that as is predicted by Theorem \ref{e_sA}, $\Phi_2$ has no eigenvalue when $A$ is small. Our analytical approximation (\ref{s_Phi12}) predicts very well the appearance of the eigenvalue of $\Phi_2$.

\begin{figure}[tbh]
\begin{center}
\hspace{0.4\textwidth}
\includegraphics[width=.4\textwidth]{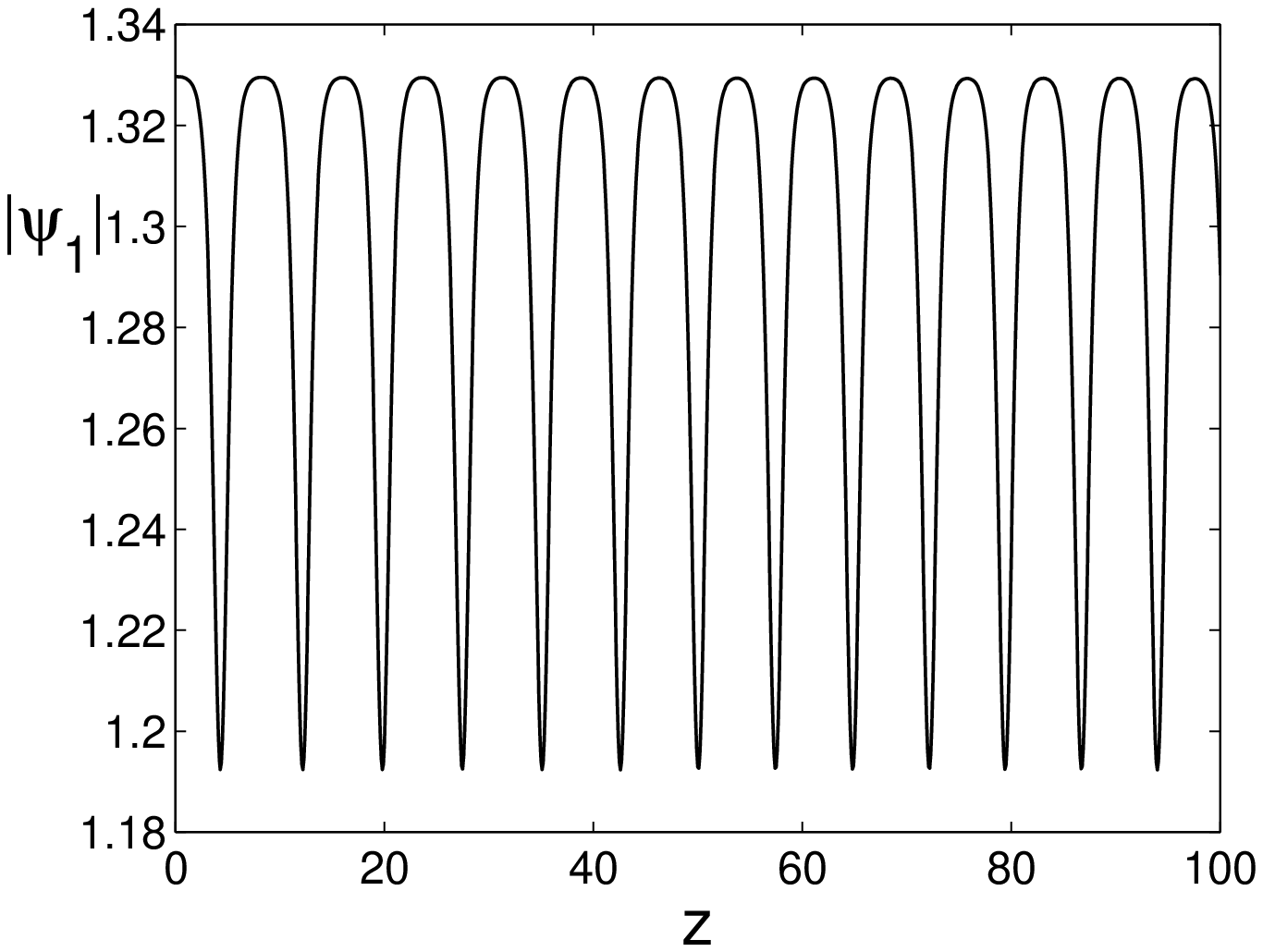}\\
\hspace{0.4\textwidth}
\includegraphics[width=.4\textwidth]{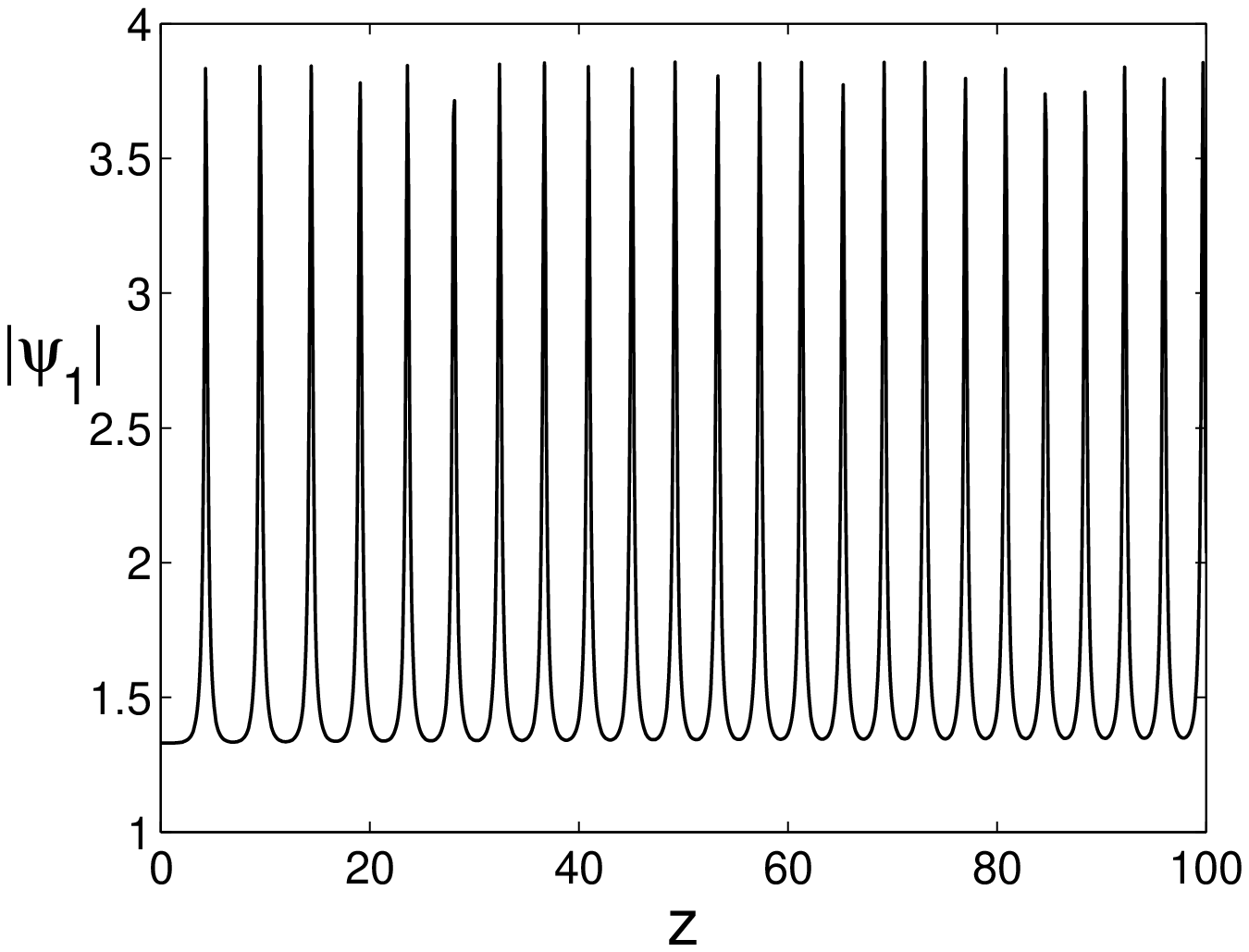}
\end{center}
\caption{The instability dynamics of $\Phi_1$ for $A=8.46$ and $\Delta=10$. It is presented that even with a tiny different perturbation the dynamics can be significantly different (see the text).}
\label{fig2}
\end{figure}

Because it is known that $\Phi_1$ is unstable in its entire existence region, it is of interests to see how the dynamics concerning the instability. In Fig.\ (\ref{fig2}) we present the evolution of $\Phi_1$ for a parameter value $A\equiv8.46$ ($A$ is already at this value from the beginning $z=0$, as opposed to Fig.\ \ref{fig1} where $A$ is 0 in the beginning and gradually increases to a constant). The top left panel presents the dynamics of $\Phi_1$ with the initial condition $\psi_n(z=0)=\Phi_1(n,A)-10^{-4}$. The initial condition $\Phi_1(n,A)$ is obtained numerically from Eq.\ (\ref{eq2}). The top right panel depicts the behavior of the first site in time where one can see that the instability manifests in the form of soliton's oscillations. Interestingly, if we start with an initial condition of the form $\psi_n(z=0)=\Phi_1(n,A)+10^{-4}$, the solution has a similar instability behavior but with a different oscillation maximum. The dynamics is presented in the bottom panels of Fig.\ \ref{fig2}. It is important to note that with such a small change, the dynamics can be significantly different. This duality therefore might be employed as a small intensity light detector similar to the proposal of \cite{khom05}.

\begin{figure}[tbh]
\begin{center}
\includegraphics[width=.4\textwidth]{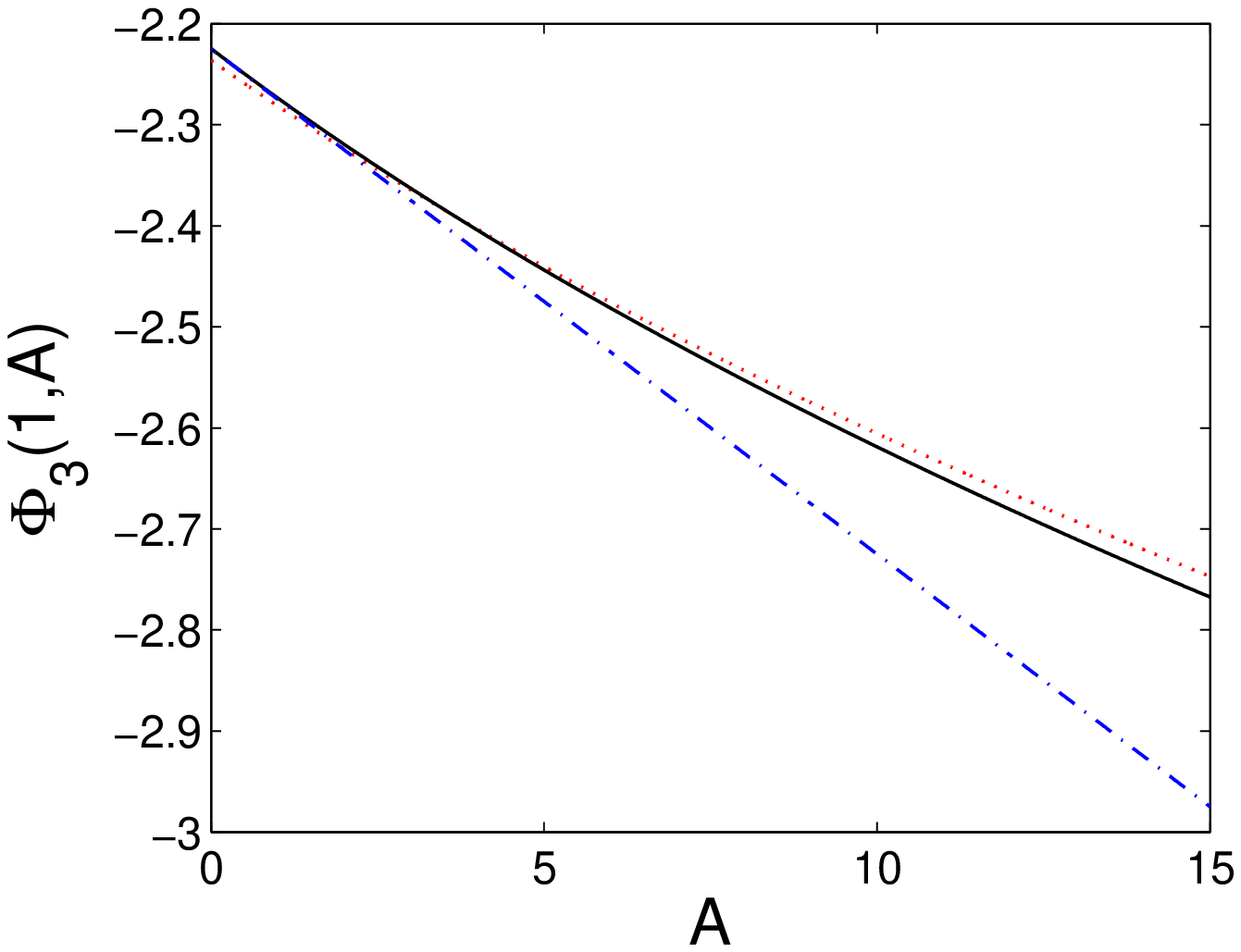}
\hspace{0.4\textwidth}
\includegraphics[width=.4\textwidth]{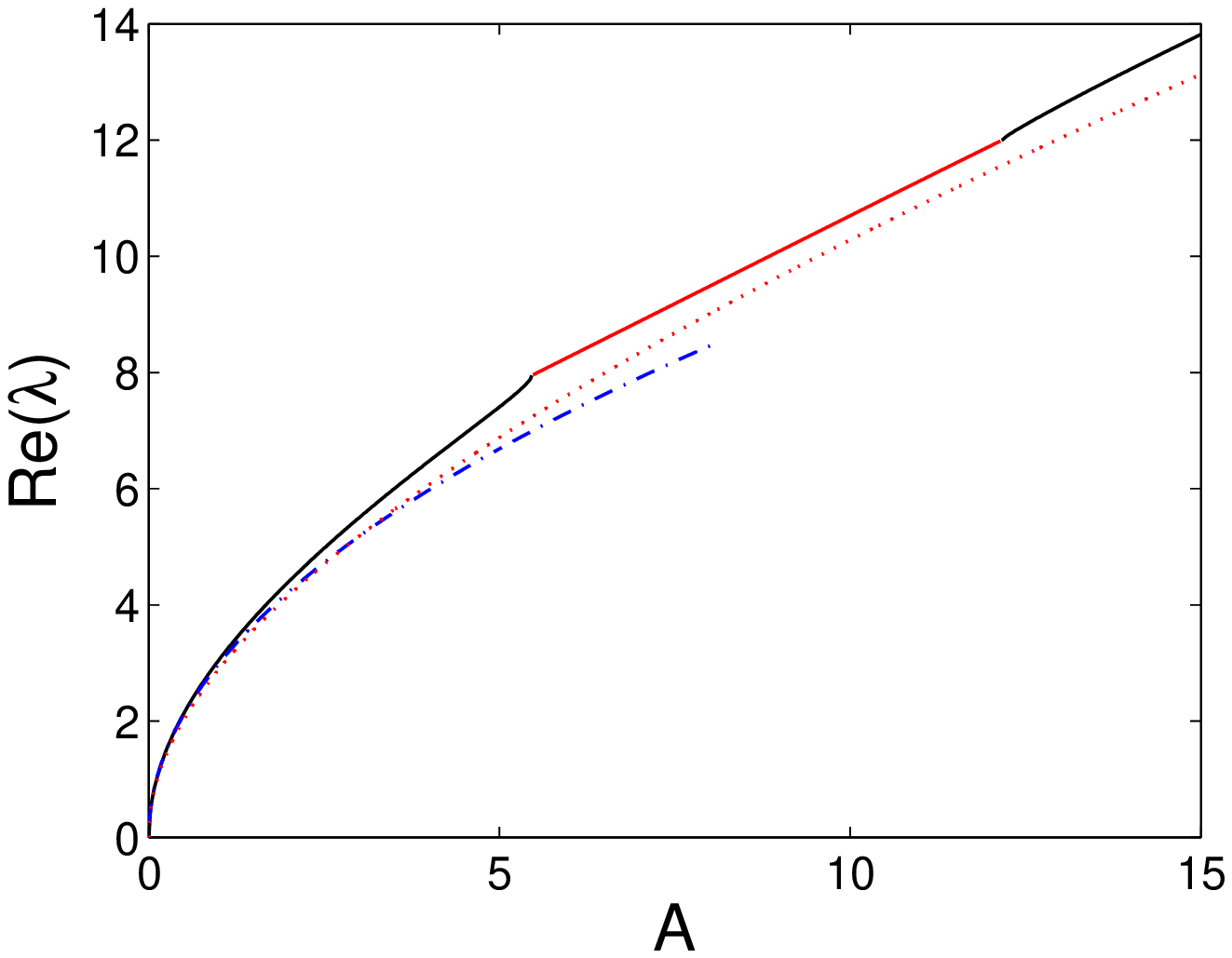}
\includegraphics[width=.4\textwidth]{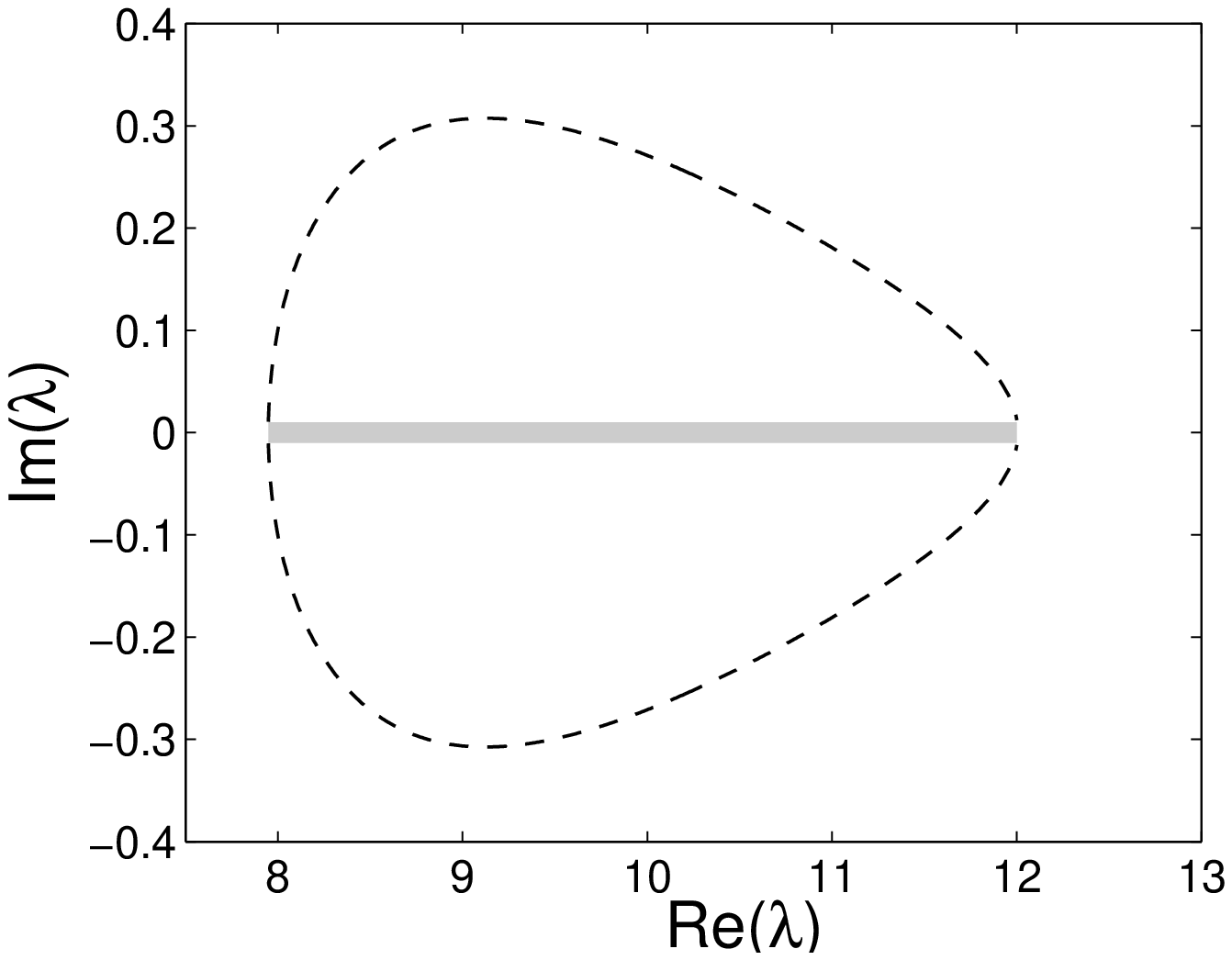}
\end{center}
\caption{Similar to Fig.\ \ref{Ath} but for the soliton $\Phi_3$. Top left panel shows the numerical results for the existence of $\Phi_3$ vs.\ the driving amplitude $A$ (solid line). Presented is the value of the solution at the first site, i.e.\ $\Phi_3(1,A)$. Bottom left panel presents the stability of the soliton. The red solid line that separates the black solid line indicates that the soliton is unstable in this region. The behavior of the critical eigenvalue in the complex plane is depicted in the bottom right panel. In the panel, the parametric variable is the driving parameter $A$. The top right panel shows the dynamics of the soliton when it is unstable. See the text for the analytical approximation curves.}
\label{fig3}
\end{figure}

We have analyzed as well numerically the existence and the stability of the soliton $\Phi_3$. We summarize our results in Fig.\ \ref{fig3}. The numerical result for the existence of the soliton is shown in the top left panel of the figure. Our analytical approximations (\ref{Phi3}) and (\ref{PhijAg}) are shown in dash-dotted and dotted lines, respectively, where one can see the good agreement between the numerical and the analytical result.

After studying the existence of the discrete soliton, we next present our stability analysis of the soliton. Shown in the bottom panels of Fig.\ \ref{fig3} is the numerically obtained critical eigenvalue of $\Phi_3$ as a function of $A$. In the bottom left panel is the real part of the critical eigenvalue. It is clear that when $A=0$, the eigenvalue is a double eigenvalue at zero. As soon as $A$ is increased, the zero eigenvalue bifurcates along the real line. At a critical driving amplitude, the eigenvalue collides with the lower boundary of the continuous spectrum. The result of the collision is the bifurcation of the eigenvalue into the complex plane resulting in an eigenvalue with nonzero imaginary part. In the bottom right panel, we present the trajectory of the eigenvalue in the complex plane as $A$ is increased. One can then see that there is also another critical amplitude above which the eigenvalue becomes real again, i.e.\ the soliton becomes stable. In the region where the imaginary part is nonzero, we depict the curve in the bottom left panel of Fig.\ \ref{fig3} in solid red line. We also compare it with our analytical approximations Eqs.\ (\ref{sta1}) and (\ref{s_Phi0}), that are shown in dash-dotted and dotted line, respectively. In Theorem \ref{s_bA}, it is stated that our analytical approximation is inconclusive for the case of $A$ large. As can be seen from Fig.\ \ref{fig3}, our analytical approximation Eq.\ (\ref{s_Phi0}) is always real. It is because when the real part of the eigenvalue is in the region of the continuous spectrum, our assumption that the eigenfunction is fastly decreasing is not justified. 

It is then interesting to see the dynamics of the instability. In the top right panel, we depict the evolution of an unstable discrete soliton of type $\Phi_3$. The parameter values are depicted in the figure. The setup for the driving amplitude is similar to the setup of Fig.\ \ref{fig2}.

Regarding the involvement of $\Phi_3$ in the dynamics of the driven boundary waveguides (\ref{eq1}) (see Fig.\ \ref{fig1}), it is not clear whether when $\Phi_2$ disappears it evolves into $\Phi_3$.

\section{Conclusions}
We have analyzed mathematically the mechanism of supratransmissions observed in a boundary driven discrete nonlinear Schr\"odinger equation describing electromagnetic fields in waveguide arrays. We have shown that the source of the phenomenon is the presence of a saddle-node bifurcation between a stable discrete soliton and an unstable one. We have shown as well numerically that the unstable one can exhibit a different dynamics, sensitive to the perturbation. We therefore argue that it might be possible to propose it as a weak signal light detector.

\section*{Acknowledgments}
The author wishes to thank Panayotis Kevrekidis and Ramaz Khomeriki for numerous useful interactions and discussions. The manuscript has greatly benefited from the constructive comments and suggestions of two anonymous reviewers.


\end{document}